\documentclass[aps,prl,nofootinbib,twocolumn,superscriptaddress]{revtex4}

\usepackage{graphicx}
\usepackage{amsmath}
\usepackage{slashed}
\usepackage{color}
\usepackage[hidelinks]{hyperref}

\newcommand{\be}{\begin{equation}}
\newcommand{\ee}{\end{equation}}
\newcommand{\br}{\begin{eqnarray}}
\newcommand{\bea}{\begin{eqnarray}}
\newcommand{\eea}{\end{eqnarray}}
\newcommand{\er}{\end{eqnarray}}
\newcommand{\ba}{\begin{array}}
\newcommand{\ea}{\end{array}}
\newcommand{\bi}{\begin{itemize}}
\newcommand{\ei}{\end{itemize}}
\newcommand{\bn}{\begin{enumerate}}
\newcommand{\en}{\end{enumerate}}
\newcommand{\bc}{\begin{center}}
\newcommand{\ec}{\end{center}}

\newcommand{\beq}{\begin{equation}}
\newcommand{\eeq}{\end{equation}}

\newcommand{\GeV}{{\rm GeV}}

\newcommand{\gsim}{\lower.7ex\hbox{$\;\stackrel{\textstyle>}{\sim}\;$}}
\newcommand{\lsim}{\lower.7ex\hbox{$\;\stackrel{\textstyle<}{\sim}\;$}}

\begin{document}

\title{Dark Matter Self-Interactions via Collisionless Shocks in Cluster Mergers}

\author{Matti Heikinheimo}
\affiliation{National Institute of Chemical Physics and Biophysics, R\"avala 10, 10143 Tallinn, Estonia}

\author{Martti Raidal}
\affiliation{National Institute of Chemical Physics and Biophysics, R\"avala 10, 10143 Tallinn, Estonia}
\affiliation{Institute of Physics, University of Tartu, Estonia}

\author{Christian Spethmann}
\affiliation{National Institute of Chemical Physics and Biophysics, R\"avala 10, 10143 Tallinn, Estonia}

\author{Hardi Veerm\"ae}
\affiliation{National Institute of Chemical Physics and Biophysics, R\"avala 10, 10143 Tallinn, Estonia}
\affiliation{Institute of Physics, University of Tartu, Estonia}

\date{\today}

\begin{abstract}
While dark matter self-interactions may solve several problems with structure formation, so far only the effects of two-body scatterings of dark matter particles have been considered. We show that, if a subdominant component of dark matter is charged under an unbroken $U(1)$ gauge group, collective dark plasma effects need to be taken into account to understand its dynamics. Plasma instabilities can lead to collisionless dark matter shocks in galaxy cluster mergers which might have been already observed in the Abell 3827 and 520 clusters. As a concrete model we propose a thermally produced dark pair plasma of vectorlike fermions. In this scenario the interacting dark matter component is expected to be separated from the stars and the non-interacting dark matter halos in cluster collisions. In addition, the missing satellite problem is softened, while constraints from all other astrophysical and cosmological observations are avoided. 
\end{abstract}

\maketitle

\section{Introduction}

Dark matter (DM) contributes a significant fraction to the energy density of the Universe, with an abundance about five times as large as that of baryonic matter. However, our knowledge about this prevalent form of mass in the Universe is quite limited. The dark sector could very well consist of several distinct species of particles, each with their own interactions and dynamics.

Observations of the Bullet Cluster (1E 0657-558) and galactic DM halos suggest that most of DM is collisionless~\cite{Markevitch:2003at, Randall:2007ph, Peter:2012jh}. Nevertheless, there are hints that at least a subdominant component of DM might be self-interacting. Apart from the missing satellite and cusp vs.~core problems, some observations of galaxy clusters support this claim. A recent weak lensing study of the Abell 3827 cluster \cite{Massey:2015oza} discovered an offset between the distribution of DM and the visible stars in the central galaxies of the cluster, which can be interpreted as a signature of DM self-interactions. Similarly, the distribution of DM in the Abell 520 cluster has been reconstructed from weak gravitational lensing \cite{Mahdavi:2007yp, Jee:2014hja}, and it has been suggested that a mass peak coinciding with the visible hot gas is required.

 It is difficult to conceive how these observations could be explained with collisionless DM; and in the case of Abell 520 assuming that all of dark matter is self interacting does not result in a correct DM distribution either~\cite{Kahlhoefer:2013dca}. Instead, an explanation might require that a subdominant component of DM behaves similarly to the visible ionized gas---developing shock fronts while its kinetic energy is converted into heat. In a cluster collision this interacting component of DM will then be subject to dissipation and consequently slow down, analogously to the X-ray emitting hydrogen gas, thus explaining the excess of dark mass on top of the visible gas in Abell 520.

The possible observation of DM plasma is compatible with recently proposed models of a dark disk within our own Galaxy \cite{Fan:2013yva}. In these models, the galactic dark plasma collapses into a thin disk through radiative cooling. In order for this collapse to occur within the lifetime of the galaxy, a light ``dark electron'' is required, which is, however, not necessary for generating the mass distribution observed in the Abell 3827 and 520 clusters.

As a minimal model capturing the essential features of plasma dynamics, we study a specific form of DM, a dark pair plasma consisting of vector-like fermions and anti-fermions that interact via dark photons---gauge bosons of an unbroken dark $U(1)$ gauge symmetry. This model can explain the existence of starless DM halos in galaxy cluster mergers, as well as solve some well known structure formation problems, while it is only weakly constrained by Cosmic Microwave Background (CMB) measurements and Big Bang Nucleosynthesis (BBN). Effects of energy dissipation in the dark sector have been considered previously \cite{Foot:2014uba}, specifically in the context of mirror dark matter \cite{Silagadze:2008fa,Foot:2013nea,Foot:2014mia}. Some of the earlier studies of $U(1)$ charged DM \cite{Ackerman:mha,Feng:2009mn,McDermott:2010pa,Holdom:1985ag,Baek:2013qwa,Baek:2013dwa} have considered Debye shielding in the dark plasma. The possible significance of the Weibel instability in galaxy collisions was mentioned in \cite{Ackerman:mha}. Recent advances in the physics of pair plasmas \cite{Bret:2009, Bret:2013qva, Bret:2014ufa} allow us draw a more definitive picture in this work. 

Up to now, studies of DM self-interactions in cluster collisions have mainly focused on the effects of individual scattering events. The aim of the present work is to point out that two-body scattering is negligible compared to the collective plasma effects, and that plasma physics should be the starting point for understanding the phenomenology of charged dark matter. We argue that models where all of DM is charged are ruled out by observations of cluster collisions unless the charge is extremely small. By keeping this in mind we propose a scenario where only a subcomponent of DM is charged.

A dark pair plasma is an appealing form of DM both theoretically as well as experimentally. We show that it can be a thermal relic, avoiding the complicated production mechanisms required for asymmetric DM such as atomic DM \cite{Kaplan:2009de,Kaplan:2011yj,Cline:2012is,CyrRacine:2012fz,Fischler:2014jda}. It does not generate dark acoustic oscillations (DAO) at observable scales, is consistent with CMB measurements, and its dark halos do not collapse to disks, due to inefficient dark bremsstrahlung. In addition, dark $U(1)$ models may solve one of the major unexplained puzzles of the standard model (SM)---the hierarchy of Yukawa couplings that spread over at least 6 orders of magnitude \cite{Gabrielli:2013jka}.

\section{Dark Plasma Dynamics}

\subsection{A Minimal Model}

Plasma is a state of matter in which collective effects, mediated by long range interactions, dominate over hard, short-range collisions of particles. The minimal model of a dark plasma is a pair plasma consisting of thermally produced vector-like dark fermions and anti-fermions. The Lagrangian of the dark sector is dark QED:
\begin{align}
		\mathcal{L}
	= \mathcal{L}_{SM} - \frac{1}{4} F_{D\mu\nu}F^{\mu\nu}_{D} + \bar{\chi}\left(i \slashed{D}  - m_{D}\right)\chi,
\end{align}
where $D_\mu = \partial_\mu - e_{D} A_\mu^{(D)}$ is the covariant derivative, $F_D^{\mu\nu}$ is the field tensor of the dark photon $A^{\mu}_{D}$, $\chi$ is the interacting DM component with mass $m_{D}$ and $e_{D}$ is the dark $U(1)$ charge.  

This Lagrangian can be extended by including a kinetic mixing term $F_{D\mu \nu} F^{\mu \nu}$ that results in electrically charged DM. Such a term is severely constrained by recombination and halo dynamics \cite{McDermott:2010pa}. Therefore, based on experimental results, we assume it to be negligible for the rest of this work.\footnote{To avoid generating the problematic kinetic mixing term radiatively \cite{Holdom:1985ag}, we have to assume that particles charged under both the hidden $U(1)$ and the SM hypercharge do not exist, or that they only exist in complete non-degenerate multiplets of a unifying gauge group. In this case, setting the tree-level kinetic mixing term to zero does not require any finetuning.}

Nevertheless, we require an interaction between the two sectors at some high scale to bring them into thermal equilibrium. It is in principle enough to just assume that the visible and hidden sectors are connected by their coupling to the inflaton, and therefore share a common reheating temperature. Here we will omit the details of this interaction, since they are irrelevant to the purpose of this paper. Instead we refer to \cite{Fan:2013yva} where various possibilities for generating the coupling between the visible and hidden sectors without inducing the kinetic mixing term are reviewed. 

We then assume that the dark sector is populated in the early Universe through freeze-out of some feeble interaction with the SM above the electroweak scale. Therefore the temperatures of the two sectors coincide at this time, but will evolve differently as the relativistic degrees of freedom drop out of equilibrium in the two sectors.The ratio of the dark sector temperature $T_{D}$ to the photon temperature $T_{\gamma}$ is fixed by entropy conservation
\begin{align}\label{eq:TD/T}
	\zeta
	\equiv	\frac{T_{D}}{T_{\gamma}}
	= 	\left(\frac{g_{*s,\gamma}(T_{\gamma})/g_{*s,\gamma}(T_{*})}{g_{*s,D}(T_{D})/g_{*s,D}(T_{*})}\right)^{1/3},
\end{align}
where $g_{*s,D}$ and $g_{*s,\gamma}$ are the numbers of relativistic degrees of freedom in the two sectors, and $T_{*}$ is the temperature at which the dark and visible sectors were presumably in thermal equilibrium.

The relic density of the fermionic interacting DM is fixed through freeze-out of the annihilation into dark photons. The thermally averaged cross section for the process $\chi \chi \rightarrow \gamma_D \gamma_D$ in the limit $T_{D}\ll m_D$ is
\begin{align}
	 \langle \sigma v \rangle 
	 = \frac{\alpha_D^2\pi}{m_D^2}
	 + \mathcal{O}\left(\frac{T_D}{m_D}\right),
\label{eq:cross section}
\end{align}
where $\alpha_D = e_{D}^{2}/4\pi$ is the fine structure constant of the dark $U(1)$. The Sommerfeld enhancement of this cross section has a negligible impact on the abundance of the dark fermions \cite{Feng:2009mn} and will be ignored. By solving the Boltzmann equation (using the procedure described e.g. in \cite{Gondolo:1990dk}) in terms of the dark sector temperature and expressing the final result in terms of the temperature of the visible sector we can estimate the relic abundance of the interacting DM component as
\begin{align}
	\Omega_{\chi} h^2 
	\approx0.3 \frac{\sqrt{g_{*}}}{g_{*s,\gamma}}
	\left(\frac{m_D}{100 \GeV}\right)^{2} 
	\left(\frac{\alpha_{EM}}{\alpha_{D}}\right)^{2} 
	\left(\frac{x_f \zeta_f}{25}\right),
\end{align}
where $\alpha_{EM} = 1/137$, $x = m_D/T_{D}$ and $g_{*}$ is the effective number of relativistic degrees of freedom in both sectors, evaluated at the time of freeze out. The freeze-out temperature can be approximated by
\begin{align}
	x_f \approx  26 + \ln\left(\left(\frac{100}{g_{*}}\right)^{1/2}\left(\frac{100 \GeV}{m_D}\right)\left(\frac{\alpha_{D}}{\alpha_{EM}}\right)^{2}\right)
\end{align}
and the ratio of the hidden and visible sector temperatures at freeze out assumes values in the range \mbox{$\zeta_f \in (0.5,1.5)$}, depending on the fermion mass.

We set this relic abundance to $\Omega_{\chi} = \xi\,\Omega_{\mathrm{CDM}}$, where $\xi$ is the fraction of the interacting species and \mbox{$h^2 \Omega_{\mathrm{CDM}} = 0.1198\pm0.0015$} \cite{Planck:2015xua} is the overall DM abundance. Thus, for $\xi=30$\% interacting DM, the expected relic abundance is obtained approximately if
\begin{align}
	\alpha_D \approx 10^{-4} \frac{m_D}{{\rm GeV}}.
\end{align}

\subsection{Collisionless Shocks}

Collisionless shocks are a prevalent phenomenon in astrophysical plasmas \cite{Bret:2015qia}, and have been observed e.g.~in the Earth's bow shock, in the expansion of supernova remnants into the interstellar medium, and in the behavior of the ionized gas in galaxy collisions. In all those situations, the mean free path of particles in the plasmas is orders of magnitude larger than the physical size of the shock fronts. 
  
The physics of collisionless shocks, which arise from collective plasma instabilities, is an active research topic. Laboratory experiments and computer simulations have investigated the formation of instabilities in relativistic and non-relativistic plasmas consisting of electrons and protons, and of electron/positron pairs. 

The formation of collisionless shocks can be roughly divided into two phases \cite{Bret:2013qva}. The first phase consists of the the buildup of the instabilities, followed by their saturation. The buildup phase can be studied by linear approximations and is relatively well understood. Depending on the type of plasma different instability modes can dominate~\cite{Bret:2009}.

In the counter-streaming situation that is relevant in cluster mergers, initial fluctuations in the counter-streaming electric currents in the plasma give rise to magnetic fields. These fields enhance the electric currents, which in turn enhance the magnetic fields, leading to an exponential growth.
Finally the field strength will be large enough to stop the stream of incoming particles, and a shock front develops.

The second phase, where the electromagnetic fields and currents reach saturation, is highly nonlinear and can be studied only by means of numerical simulations. Most of the dissipation of kinetic energy takes place in this phase. The latter can be roughly understood as an effect caused by scattering of the upstream particles from the strong electric and/or magnetic fields generated during the first phase.

The effect of the collisionless shocks in a cluster merger can be understood as follows. The DM halos of the colliding subclusters are initially in a stable equilibrium state. Once they begin to overlap, an unstable counter-streaming situation arises, and a shock front quickly develops as described above. This shock front then propagates through the subcluster halo, heating and slowing down the interacting component of DM, similarly to what is seen in the X-ray emitting visible gas. Consequently, a generic prediction of the dark plasma model is the existence of the bow-shaped dark matter shock fronts, that should be visible in the weak lensing reconstructions of the cluster merger events, given sufficient angular resolution.

Our suggestion is that these effects might have already been discovered in the Abell 520 and 3827 clusters. In Abell 520 an excess of DM on top of the visible X-ray emitting plasma between the subclusters is observed. This can be interpreted as the interacting component of dark matter that was slowed down due to the shocks, similarly to the visible plasma. 

In Abell 3827 the separation between the stars and the center of mass of the dark matter in the central galaxies can be interpreted along the lines of \cite{Kahlhoefer:2015vua}: The interacting component of DM in the galaxies is counter-streaming against the DM in the main cluster halo. The resulting shocks create an effective drag force slowing down the interacting DM component of the galaxies, resulting in a separation between this component and the rest of the mass. Therefore, the center of mass of the total dark matter distribution is separated from the stars, even though the main component of DM remains on top of the stars in this scenario. Given a high enough resolution, this effect should be observable as separated starless dark matter clumps, similar to what is observed in a larger scale in Abell 520.

\subsection{Shock Formation Time Scale}

To show that these effects indeed are relevant in a typical cluster merger, we shall now examine the fundamental characteristics of the dark plasma---its Debye length, the plasma parameter and the plasma frequency. For the colliding intracluster DM halos, we set the size to $R=200$ kpc and the mass to $M=4 \cdot 10^{13} M_\odot$, corresponding to the dimensions of the colliding Abell 520 subclusters as determined from weak lensing data \cite{Jee:2014hja}. Assuming uniform distribution, the average density of the interacting DM is then $1.36 \cdot 10^{-2}$ GeV/cm$^3$. 

This information, together with the interaction strength $\alpha_D$, is sufficient to calculate the time scale for the plasma instabilities. The mean free path of DM particles  depends on the temperature of the self-interacting DM component, which we can estimate from the virial theorem for each of the colliding sub-clusters 
\begin{equation}
      T_{\mathrm{vir}} = \frac{G_N M m_D}{n_{\mathrm{dof}} R} = \frac{M m_D}{3 m_{P}^2 R} = 3.2 \cdot 10^{-6} \, m_D.
\label{eq:dm_temperature}
\end{equation}
Here $G_N=1/m_{P}^2$ is Newton's constant, and $n_{\mathrm{dof}} =3 $ is the number of degrees of freedom of a single particle.

Returning to the dark plasma, the Debye length in our minimal model is 
\begin{equation}
	\lambda_D 
	= \sqrt{\frac{T_{\rm vir}}{4 \pi \alpha n}} 
	\approx \, 30.7 \mbox{ km }  \sqrt{\frac{m_{D}}{\mbox{GeV}}} .
\end{equation}
If two particles are separated by more than this distance, the Coulomb interaction between them is effectively shielded by free charge carriers. Clearly this distance is diminutive in comparison with astrophysical scale.

The plasma parameter $\Lambda$ is defined as the number of charge carriers within a sphere of radius $\lambda_D$,
\begin{equation}
	\Lambda 
	= \frac{4 \pi}{3} \; \lambda_D^3 n
	\approx 1.7 \cdot 10^{18} \, \sqrt{\frac{m_{D}}{\mbox{GeV}}} .
\end{equation}
Since $\Lambda \gg 1$, the the plasma is weakly coupled and collective effects caused by long range forces dominate plasma dynamics. The characteristic time scale for collective plasma effects is the inverse of plasma frequency 
\begin{equation}
	1/\omega_{p} 
	= \left( \frac{m_D}{4 \pi \alpha n} \right)^{1/2}
	\approx 57.2 \; \mbox{ms} \; \sqrt{\frac{m_D}{\mbox{GeV}}},
\end{equation}
where $n$ is the number density and $m_D$ the mass of the DM particles.

The mean free path of charged particles in plasma is of the order of \cite{Kulsrud:2004}
\begin{equation}
	\lambda_{\mbox{\scriptsize mfp}} 
	= \lambda_D \; \frac{\Lambda}{\log \Lambda}
	\approx 39.4 \mbox{ kpc} \, \left( \frac{m_{D}}{\mbox{GeV}} \right) .
\end{equation}

The time it takes to form a collisionless shockwave can be estimated by considering the instability growth rate. In a symmetric non-relativistic collision of two cold counter-streaming pair plasmas the dominant instability mode is the Two-Stream mode for which the instability growth rate is of the order of the plasma frequency \cite{Bret:2009}. A realistic description of the collision certainly requires a more complicated set-up than a simple cold counter-streaming plasma as the latter does not account for the possible inhomogeneities and temperature or velocity dispersion inside the cluster. However, the latter considerations should not significantly affect the order of magnitude estimate given here. A conservative order of magnitude estimate of the time scale of shockwave formation is then
\begin{align}\label{eq:c_shock_t}
	\tau_{s} 
	\approx 10^{3} \omega_{p}^{-1}
	\approx 57.2 \; \mbox{s} \; \sqrt{\frac{m_D}{\mbox{GeV}}} .
\end{align}

The distance for which plasma instabilities become relevant can be estimated by multiplying our estimate for the shock formation time in eq.~(\ref{eq:c_shock_t}) by the typical speed of a cluster collision,
\begin{equation}
   \lambda_s \approx \tau_s v_{\mathrm{col}} \sim 10^5 \ {\rm km},
\end{equation}
which is certainly much smaller than the mean free path.  Thus, the plasma instabilities affect the dynamics of the interacting DM component at time and distance scales much shorter than the two-body scattering processes. This leads us to conclude that DM charged under an unbroken $U(1)$ interaction should be treated primarily as a plasma that develops collisionless shocks at relatively small scales. Consequently, at larger scales it behaves effectively as a collisional fluid, even if the two-body scattering rate seems to suggest that collisions are insignificant.

We are not aware of studies of non-relativistic pair plasmas. However, a study of relativistic pair plasma collisions \cite{Bret:2014ufa} suggests a time scale that is even an order of magnitude smaller than estimated in eq.~(\ref{eq:c_shock_t}). Our argument here is based on the instability growth rates in the theoretically well understood linear regime. A numerical study, which will provide a more conclusive treatment, is under preparation.

\subsection{Atomic Dark Plasma}

Our minimal model assumes that the interacting subcomponent of DM exists in the form of a collisionless pair plasma. It is also conceivable that the dark plasma could consist of light ``dark electrons'' and more massive ``dark protons'', thus imitating the visible sector. Such scenarios of atomic DM require a non-thermal, asymmetric production history. Here we do not discuss any such models in detail, but note that any mass non-degeneracy of the particles making up the plasma does not significantly affect the estimate of the instability growth rate \cite{Bret:2009}. Therefore the collisionless shock behavior will also dominate the dynamics of atomic dark plasmas.

An interesting possibility for explaining both the subdominant interacting DM species as well as the collisionless main component could be a model of partially ionized atomic dark matter, where the dark plasma consists of dark ions and dark electrons, while the main component of DM is composed of the neutral atoms.

\section{Observational Constraints}

\subsection{Bullet Cluster}

This work was motivated by observations of the Abell 3827 and 520 clusters, but many similar systems exist \cite{Harvey:2015hha}. Perhaps the most unambiguous of those is the Bullet Cluster, from which constraints have been derived on the interactions of DM. It has been shown that no more than 30\% of the total DM mass can be lost from the subcluster as it passes through the main halo \cite{Markevitch:2003at}.  Thus, throughout this work we will assume that the fraction of interacting dark matter species obeys \mbox{$\xi \leq 0.3 $.} This rules out models where all of dark matter is charged under an unbroken $U(1)$ as long as the coupling constant is not negligibly small.

\subsection{BBN}

Changing the properties of DM changes the dynamics of the early Universe, and therefore any deviations from the usual $\Lambda$CDM model will be strongly constrained. New relativistic degrees of freedom change the expansion history of the Universe during BBN, leading to a constraint which is usually expressed as a limit on the effective number of light neutrino species \cite{Planck:2015xua}, 
\begin{align}\label{eq:bound_Neff}
	N_{\mathrm{eff}} = 3.04 + 2 \zeta_{BBN}^{4} = 3.15 \pm 0.23,
\end{align}
Assuming one species of fermions we see that the ratio \eqref{eq:TD/T} of the temperatures of the dark and visible sectors at BBN is $\zeta_{BBN} = 0.52$, implying that $N_{\mathrm{eff}} = 3.18$. Therefore Eq.~\eqref{eq:bound_Neff} is satisfied within $1\sigma$.

By considering an extended scenario with $N_D$ effective relativistic dark fermions at $T_{*}$, the bound in eq.~\eqref{eq:bound_Neff} implies
\begin{align}\label{eq:bound2_Neff}
	N_D = 0.68 \pm 1.67\,.
\end{align}
Note that if all fermion masses exceed $T_{*}$, then $N_D = 0$ and eq.~\eqref{eq:bound_Neff} is always satisfied.

\subsection{CMB}

Dark matter is usually considered to be pressureless, so that all primordial DM density fluctuations start to grow immediately after they enter the horizon. If a subcomponent of DM interacts with massless dark photons, the growth of structure is suppressed until the DM and dark photons kinetically decouple. This effect could be observed in the Cosmic Microwave Background as a suppression of fluctuations at large multipole moments. 

The kinetic decoupling of the dark fermions and the dark photons occurs when the Compton scattering rate in the dark plasma drops below the Hubble rate \cite{Bringmann:2006mu}. The Compton scattering rate for the dark plasma is 
\be
	\Gamma_C = \frac{64\pi^3\alpha_D^2T_D^4}{135 m_D^3},
\ee
and the Hubble rate in the radiation dominated epoch is
\be
	H=\sqrt{\frac{4\pi^3}{45}g_*}\frac{T_\gamma^2}{m_P},
\ee
where $g_*$ is the effective number of relativistic degrees of freedom in the visible sector, with $g_*=3.36$ at temperatures well below the electron mass, and we have neglected the subdominant contribution of the colder dark sector. Setting $\Gamma_C=H$ we obtain the temperature of kinetic decoupling 
\be
	T_{\rm kin} = \left(\frac{4 g_*}{45\pi^3}\right)^\frac14\frac{\sqrt{135}}{8\zeta_{\rm kin}^2}\frac{m_D^\frac32}{\sqrt{m_P}\alpha_D},
\ee
where $\zeta_{\rm kin}$ is the ratio of the dark and visible photon temperatures at the time of kinetic decoupling. The determination of the exact effect of the DM/dark radiation coupling on the CMB is beyond the scope of this letter. Here we will simply require that decoupling happens above \mbox{$T_{\rm kin} > 640$ eV,} so that the DM/dark radiation coupling only affects multipoles above $l>2500$ and is thus unconstrained by the Planck data. This will lead to a more conservative limit than what would be allowed by a more detailed analysis, but will be used here as a robust constraint.

It should be noted that values near the lower end of this limit, or slightly below it, can help to alleviate the \emph{missing satellites} problem \cite{Bringmann:2006mu,Chu:2014lja}. The cut-off on the size of the gravitationally bounded DM structures due to kinetic coupling with the dark radiation is given by \mbox{$\sim 10^{-4}(T_{\rm kin}/10\ {\rm MeV})^{-3}M_\odot$} \cite{Loeb:2005pm}, so that for \mbox{$T_{\rm kin}\approx 0.5$ keV} the cut off is at $\sim 10^9M_\odot$, as required to ease the missing satellites problem. However, in our case only a subdominant part of DM is coupled to the dark radiation and thus the cut-off will only affect the interacting fraction of DM, so that structures smaller than the cut-off will still exist, only in fewer numbers.

\begin{figure}
\begin{center}
\includegraphics[width=0.48\textwidth]{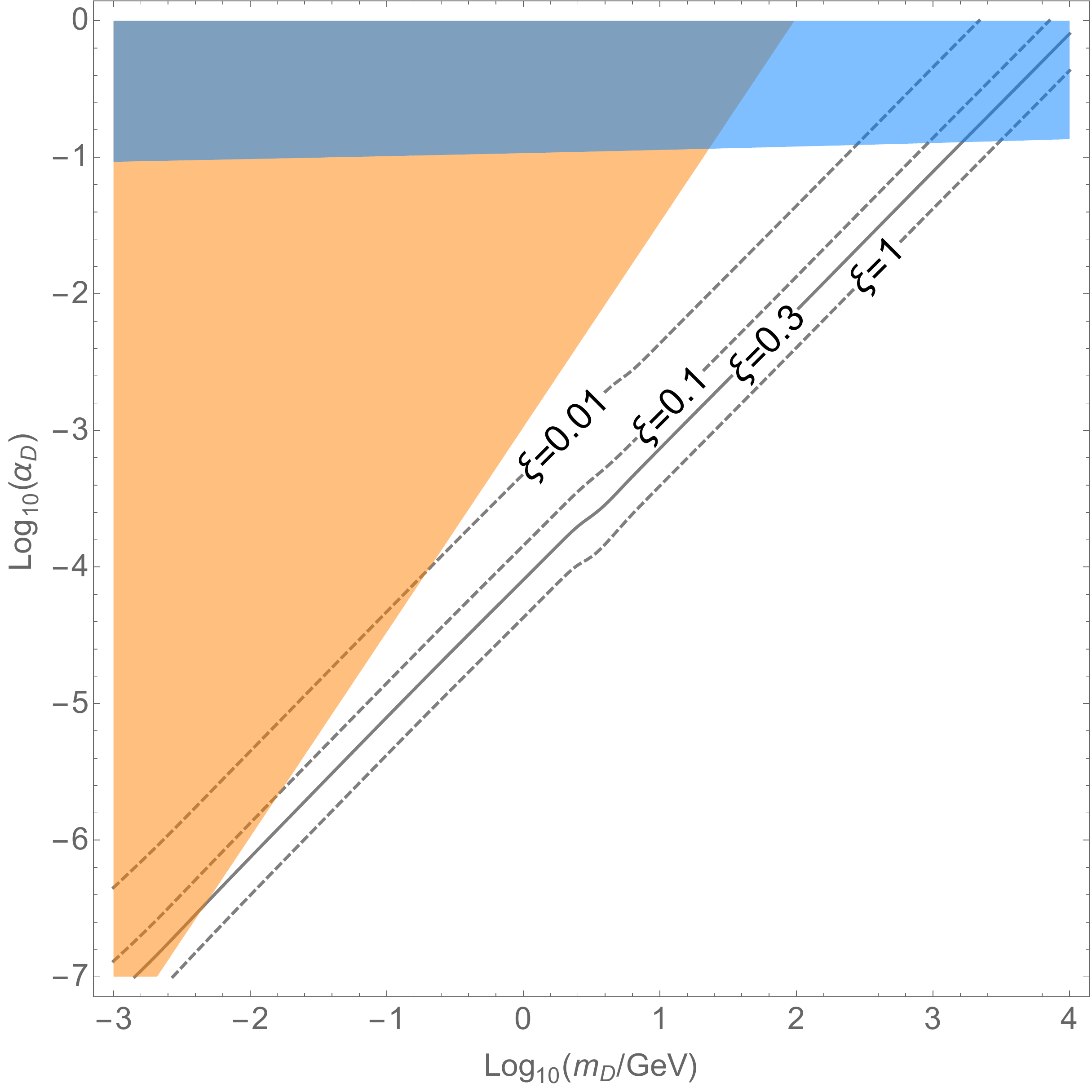}
\caption{The constraints on the parameter space of the dark pair plasma model. The black contours show the relic abundance of the dark plasma, as a fraction of the total DM abundance. The orange shaded region is disfavored by the kinetic decoupling constraint, and the lower limit of that region is favored for alleviating the missing satellites problem. The blue shaded region is excluded, in absence of a UV completion, if we require that the Landau pole of the dark $U(1)$ coupling is above the Planck scale.}
\label{fig:alpha_vs_mD}
\end{center}
\end{figure}

Figure \ref{fig:alpha_vs_mD} depicts the kinetic decoupling constraint on the $(m_D,\alpha_D)$-plane, as well as the parameter space region compatible with the relic abundance. Restricting to the region that produces the desired relic abundance, the kinetic decoupling limit gives a lower limit on the mass of the DM particle.
For $\xi = 0.3$ this limit is roughly \mbox{$m_D \gtrsim 5$ MeV}. We show also the upper limit from requiring that the Landau pole of the dark $U(1)$ lies above the Planck scale, which for $\xi = 0.3$ corresponds to roughly \mbox{$m_D \lesssim 2$ TeV}.

It should be stressed that the constraints for dark matter self interactions, including the kinetic decoupling constraint depicted in the figure, are in reality functions of $\xi$, naturally vanishing as $\xi$ goes to zero. For simplicity we only show the conservative limit, requiring that no effects are generated at observable scales in the CMB. On the other hand, due to the constraints from the Bullet Cluster discussed above, values of $\xi$ much larger than 0.3 are excluded for the complete range of $\alpha_D$ shown in the figure.

In the case of atomic DM, there is an additional constraint from the acoustic oscillation peak that results from the recombination of dark atoms \cite{Cyr-Racine:2013fsa}. This turns out to be very constraining on the parameter space of the atomic DM, but it still leaves some plausible parameter space for, \emph{e.g.}, the model proposed in \cite{Fan:2013yva}.

\subsection{Dark Halo Stability}

The dark pair plasma will heat up and become virialized both in galactic DM halos and in galaxy cluster collisions. Similar to visible matter, it can dissipate heat through dark bremsstrahlung and through Compton scattering off dark photons. The characteristic time scale for dark bremsstrahlung cooling for the plasma parameters given above is \cite{Fan:2013yva}
\begin{equation}
 t_{\mathrm{brems}} \approx \frac{3}{16} \; \frac{m_D^{3/2} T_{\rm vir}^{1/2}}{n_D \alpha_D^3} \approx 6.7 \cdot 10^{19} \; \mbox{yr}, 
\end{equation}
where the dependence on the DM mass cancels for our model: cooling becomes more efficient for smaller DM mass, but this is offset by the smaller coupling constant required to produce the correct relic density. The characteristic time scale for dark Compton scattering is even larger and can be neglected for all reasonable values of the DM mass and coupling.

A thermally produced pair plasma can therefore not efficiently dissipate heat, so that dark pair plasma halos do not collapse to disks. This is in contrast to atomic DM, consisting of asymmetric light and heavy dark fermions. Because of the non-thermal production mechanism, the dark coupling strength is here independent of the DM density. Atomic DM with light ``dark electrons'' can therefore cool down sufficiently fast to collapse to a dark disk within the lifetime of the Universe. However, the counter-streaming instabilities in the dark plasma might give rise to nontrivial substructure within the galactic DM halo. Conclusive treatment of this issue would require numerical simulations, and is beyond the scope of this letter.

\section{Discussion and Conclusions} 

In this letter we considered the possibility that a subdominant component of DM has long-range interactions and exists as a dark pair plasma. We found that a thermally produced dark pair plasma is necessarily collisionless, but will self-interact through collective plasma effects. The possibility of forming collisionless shocks will then modify the dynamics of galaxy cluster collisions, leading to effects such as an offset of DM halos as in Abell 3827, or to an isolated DM clump as in Abell 520.

To our knowledge, no numerical simulations or experiments have so far directly investigated non-relativistic collisionless pair plasmas. Our argument is however quite general, based on instability growth rates in the linear regime. For an atomic  DM plasma which imitates the SM, the analogous behavior of visible astrophysical plasmas can be directly used as a proof that collisionless shocks should also exist in the dark sector. Nevertheless, numerical studies of non-relativistic dark pair plasmas would be desirable for a more thorough treatment.

Ultimately, collisionless shocks are an efficient form of DM self-interactions,
explaining the features observed in the Abell 3827 and 520 clusters. If a galaxy moves through an intra-cluster medium of interacting DM, a drag force between the halo DM and the background cluster dark plasma is generated through plasma instabilities, potentially distorting halo shapes or even striping galaxies of the interacting DM component. In cluster mergers bow-shaped dark matter shock fronts are expected to be observable in high resolution weak lensing studies. Thus a dark pair plasma offers spectacular signatures for its discovery, and we encourage the astrophysics community to look for these effects.

\section*{Acknowledgments} We thank Boris Z.~Deshev for bringing the features of Abell 520 to our attention, as well as Antoine Bret and Luca Marzola for useful discussions. This work was supported by grants MJD435, MTT60, IUT23-6, CERN+, and by the EU through the ERDF CoE program.

\end{document}